**The Future of Office and Administrative Support Occupations in the Era of Artificial Intelligence: A Bibliometric Analysis**

Priyadarshini R. Pennathur[1*], Valerie Boksa[2], Arunkumar Pennathur[1], Andrew Kusiak[3], Beth Livingston[4]

[1]Department of Industrial, Manufacturing and Systems Engineering, University of Texas at El Paso

[2]Park Nicollet Methodist Hospital

[3]Department of Industrial and Systems Engineering, University of Iowa

[4]Department of Management and Entrepreneurship, University of Iowa

**\*Corresponding author email: prpennathur2@utep.edu**

**Abstract**

The U.S. Bureau of Labor Statistics projects that by the year 2029, the United States will lose a million jobs in the office and administrative support occupations because technology, automation, and artificial intelligence (AI) have the potential to substitute or replace the office and administrative functions performed by office workers. Despite the potential impact AI will have on office work and the important role office workers play in the American economy, we have limited knowledge of the state of the art research in office work at the intersection of emerging artificial intelligence technologies. In this study, we conducted a bibliometric analysis of the scholarly literature at the intersection of office work and artificial intelligence. We extracted literature sources from Compendex and Scopus databases and used VOSviewer for visualizing and quantifying our bibliometric analyses. Our findings from keywords analysis indicate that office automation, humans, human-computer interaction, and artificial intelligence occurred more frequently in the scholarly literature and had high link strengths. Keyword clusters from co-occurrence analysis indicate that intelligent buildings, robotics, and the internet of things are emerging topics in the office work domain. The two clusters related to ergonomics, worker characteristics, human performance, and safety indicate the types of human factors concerns that are more widely studied in office work settings. In summary, our findings on the state-of-the-art research in office work indicate that more studies have been conducted on smart buildings, robotics, and technology development for office work, compared to studies on office workers and their professional development.

**Managerial Relevance:** Understanding how technology advances impact office work and office workers will have implications for how technology-driven and human-centered decisions will be made in the future.

1. Introduction

The U.S. Bureau of Labor Statistics projects that by the year 2029, the United States will lose a million jobs in office and administrative support occupations. These losses will occur because "technology in the form of artificial intelligence (AI) and automation will substitute or supplant functions that workers in office and administrative support occupations do [1]." Business and industry professionals concur with these predictions triggered by the developments in process automation, process monitoring, data science and analytics. Automation software companies expect their sales to grow by at least 20% [2], [3], with companies like UiPath reporting market valuations upwards of $35 billion as of 2021, and software giants like Microsoft entering this business domain[2], [4]. More recently, large language models have enabled new capabilities, including knowledge creation and management, assistance in planning and scheduling tasks, and emotion management. These technological advances demonstrate the potential for significant digital transformation of office and administrative support occupations in the near future.

Workers in the office and administrative support occupations (referred to collectively as office workers in this paper), are ubiquitous in the American workplace and serve as a critical linchpin of the American economy. Their median age is 42 years [5]. They only need a high school diploma and minimal skills to get hired, and receive minimal on-the-job training, making their mobility harder if they lose their jobs (1). Moreover, as of 2021, nearly 70% of office workers are female [5]. Given these characteristics, any AI-based transformation is likely to have a more significant impact on office workers.

Despite the large-scale impact AI is poised to have on office work, and despite the profound ramifications displacements in the office workforce can have on individuals, particularly women, on organizations, and on our economy, we do not have a sufficient understanding of the current state of the art in the scholarly literature about office work and artificial intelligence. Hence, one main objective of this paper is to discuss the current state of art and emerging trends in research on office work. Additionally, the open research problems and themes at the intersection of artificial intelligence and office work are multidisciplinary, given the human, technology, economic, policy, social, gender, and equity implications of designing AI for future office work. Hence, to understand how best AI could support office work in the future and to develop work design recommendations and convergent solutions, we will need to examine multidisciplinary perspectives including human factors engineering, computer science and AI, diversity, equity and inclusion, economics, human resource management, labor relations, and public policy. However, we do not know whether current scholarly literature has approached office work and artificial intelligence with a multidisciplinary lens. Hence, another aim for this paper is to identify whether current scholarly research has included multidisciplinary perspectives in addressing problems at the intersection of office work and AI. We address our aim through a bibliometric analysis of the scholarly literature.

A bibliometric analysis summarizes knowledge from a large body of literature to understand the "intellectual structure and emerging trends" in a study discipline [6]. Given that the initial search results in databases that publish scholarly research about a topic are typically large but helpful to broadly understand the state of a study topic, a bibliometric analysis is especially beneficial in the early stages of a search for scholarly literature. Bibliometric analysis has been gaining recognition in many fields [7], [8], [9], [10], [11], [12], especially because of its capability to handle large volumes of literature and the availability of tools such as VOSviewer [13], Gephi[14], and Citespace [15] to perform the analysis. Bibliometric analysis has been used to understand the state of the art in a research domain, identify collaboration and networking patterns among scholars in specific research areas, and map the scientific knowledge in a research domain [6]. In particular, bibliometric analysis is helpful to identify knowledge gaps in research, which can then be used to define and shape a research agenda so that novel and impactful research contributions can be made to propel a scholarly area of study forward.

In this paper, we discuss findings from a bibliometric analysis of the literature to provide a macroscopic view of the core research areas in office work, emerging trends and patterns in the state of the art in office work and AI, and the multidisciplinary research and study disciplines that have been involved in investigating office work.

## 2. Methods

### 2.1 Search Strategy

#### 2.1.1 Development of search terms for scholarly literature databases

In consultation with an engineering librarian and based on several discussions within our research team, we devised inclusion criteria for the literature we desired to search for conducting the bibliometric analyses. We also developed and refined the specific search criteria we would use for searching the literature. Our main inclusion criteria were to have studies related to office workers, office work, office automation, artificial intelligence in office work, cognitive work, articulation work, well-being, training, labor and human resources, equity, and gender. We designed our search criteria such that the resulting studies focused on office work and either automation or artificial intelligence related to office work. We also focused on studies that had a direct or indirect focus on administrative support or office workers in any industry - studies either focusing on their work, or studies to develop or evaluate technologies related to office work. We focused only on publications written in English. The research team also provided input on databases that might be relevant and contain their discipline's scholarly literature. Based on this input from the research team, we then selected two major scientific literature databases to conduct our literature search: Compendex and Scopus.

First, in consultation with the engineering librarian, we decided on two broad concepts as the focus of this literature analysis: (1) Artificial Intelligence and Automation and (2) Office Workers and Office Work. Then, we provided a set of initial seed references to the engineering librarian to clarify the focus of our analysis. The engineering librarian used this seed reference list as a base to investigate how these articles were indexed in Compendex and Scopus and then harvested the terms that might reflect artificial intelligence, automation, office work, and office workers. This process resulted in an initial set of search terms.

We also approached the literature search process with a convergent and multidisciplinary lens. The research team members brought to the search process expertise in human factors engineering, data science, labor relations and human resources, and gender and equity. Each research team member with specific disciplinary expertise was asked to provide search words that fit under these two concepts and those that they consider relevant to their sub-discipline. For example, the researcher with expertise in data science provided search terms such as explainable artificial intelligence and service robotics; the researcher with human resources and gender expertise provided terms such as human resources management, labor relations, turnover intentions, gender, and equity; the research team member with human factors and training expertise provided search terms such as skill acquisition, technological skills training, human supervisory control; the research team member with cognitive engineering expertise provided terms such as articulation work, computer-supported cooperative work, etc.

We discussed, collated, and finalized all these search terms from the initial harvesting of search terms by the engineering librarian and the input from the research team. Two authors reviewed the list of search terms generated and removed terms not relevant to the focus of this work. For example, terms such as computer analysis and employee were deemed too general, and terms such as embedded systems were not relevant to the focus of our work and were removed. We also removed any duplicates. This resulted in a total of 185 search terms.

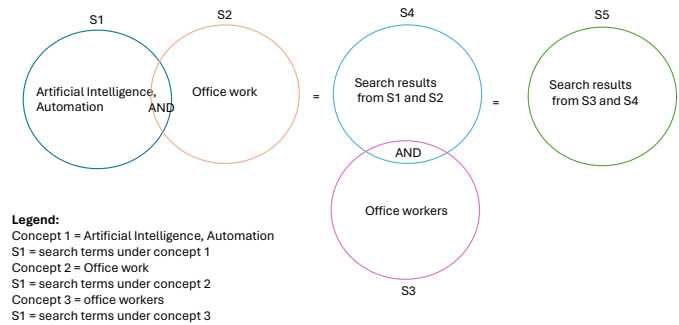

**Figure 1. Search Strategy with search concepts and search terms used in combination with Boolean Operators**

Once we finalized the search terms, we discussed an initial search strategy with the engineering librarian. We classified the search terms under the two main concepts. Then, the engineering librarian created a search query using the syntax required for the database by combining search terms under these two concepts (S1 + S2) with Boolean operators. We then further delineated the search strategy by separating the terms referring to office workers from S2 and adding a third concept for office workers and a corresponding search query, S3. This resulted in a search strategy of S1 AND S2 = S4 and then we combined S3 AND S4 = S5 (Figure 1). The engineering librarian also added controlled terms to capture the most recent papers that had not yet been assigned any controlled terms in the databases. In essence, this search query was intended to return publications at the intersection of artificial intelligence, automation, and office work, and office workers. This search strategy was used for both Compendex and Scopus. Appendix A provides the search terms used and the Boolean operators used to combine those search terms.

## 2.2 Bibliometric Analysis

Our search that combined relevant Medical Subject Heading (MeSH) terms and Boolean operators yielded 316 peer-reviewed papers in the Compendex database and 565 in the Scopus database. We performed bibliometric analysis on the search results from each database. We used VOSviewer [16], an open-source software widely used for bibliometrics, to conduct our analysis. VOSviewer allows the creation of networks based on scientific publications, authors, organizations, countries, and keywords. VOSviewer supports various types of analyses, including keyword co-occurrence analysis, citation and co-citation analysis, and co-authorship and bibliographic coupling analyses. VOSviewer generates three distinct map visualizations: a network visualization that uses different colors to depict different clusters, an overlay visualization that color codes the clusters by year, and a density visualization, where the intensity of a color indicates the significance and frequency of an item in the data.

Given our interest in exploring emerging trends in office work, we focused primarily on co-occurrence analysis, with the aim of defining a research agenda on this topic for future research. We analyzed the bibliometric data of the search results by examining the co-occurrence of keywords. We employed a full counting method, which equalized the weighting of all keyword co-occurrences. All other default parameters in VOSviewer were retained. Each keyword in the Compendex database required a minimum of five occurrences for inclusion in the analysis. Upon reviewing the keyword list, we excluded the word 'surveys', leaving a total of 71 keywords for inclusion in the analysis. The Scopus analysis required a minimum of 10 occurrences for each keyword. After reviewing the list of keywords, a thesaurus was developed to merge similar word pairs, such as 'musculoskeletal disorders' and 'musculoskeletal disorder'. Words pertaining to the study methodology rather than content were excluded (such as: 'article', 'questionnaire',

'human experiment'), leaving a total of 84 keywords in the analysis. Bibliometric visualizations of the co-occurrence analysis were generated. We also documented the keywords that occurred most frequently and their total link strengths from VOSviewer's results. Furthermore, we examined publication trends, including the publication year, the publication type, and the publication venues (top journals and conference proceedings).

3. **Study Findings**

The goal of our study was to conduct a bibliometric analysis of the scholarly literature to highlight the core research areas in office work, emerging trends and patterns to understand the state of the art in office work and AI, and the multidisciplinary areas involved in investigating office work. To achieve this goal, we conducted publication metrics analyses, trend analyses, and keyword co-occurrence analyses.

<u>3.1 Publication Metrics and Trends</u>

Figures 2 and 3 show an increase in the number of publications on office work in the 1980's and then again between 2019-2020.

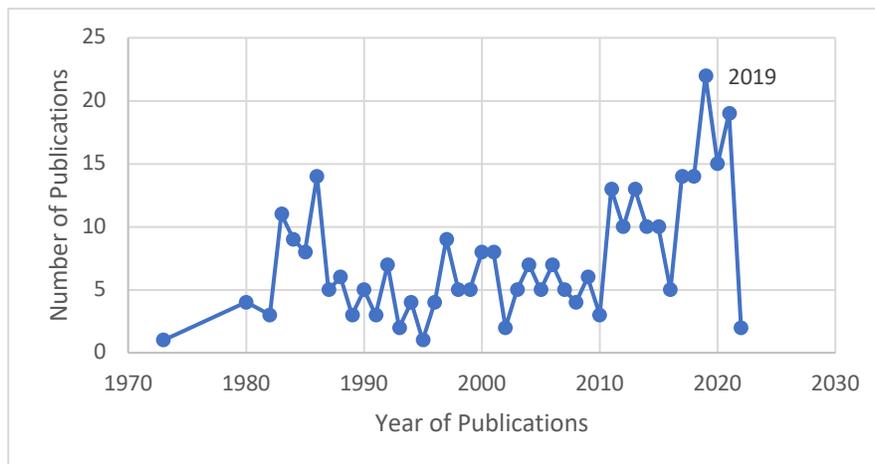

**Figure 2. Number of publications on office work from 1973 to 2023 from the Compendex Database**

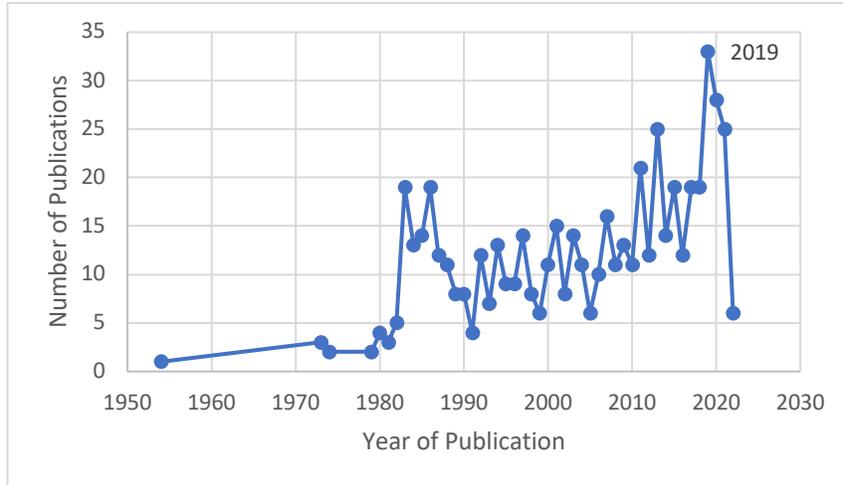

**Figure 3. Number of publications on office work from 1950 to 2023 from the Scopus Database**

Figures 4 and 5 show that, generally, conference proceedings outnumber journal publications on this topic, but the Scopus database shows slightly more journal publications.

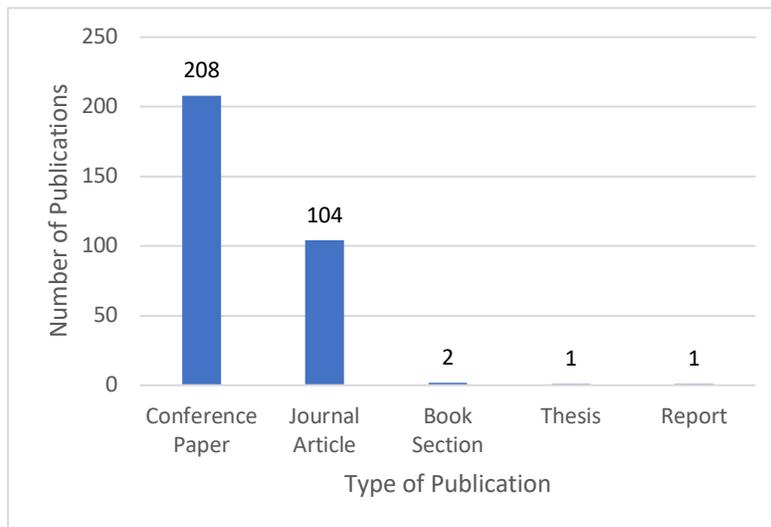

**Figure 4. Publication trends indicating the type of publications on Office Work from the Compendex Database**

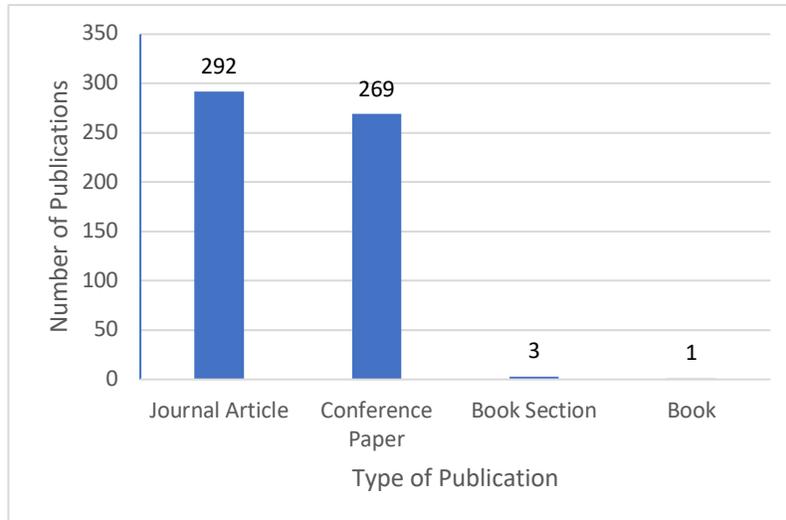

**Figure 5. Publications trends indicating the type of publications on office work from the Scopus Database**

As figures 6 and 7 illustrate, a vast majority of the research has been published in computer science journals and conference proceedings, closely followed by publications in industrial engineering, and human factors and ergonomics, both in the Compendex and Scopus databases. It is also noteworthy that some publications appear in robotics and building and environment journals.

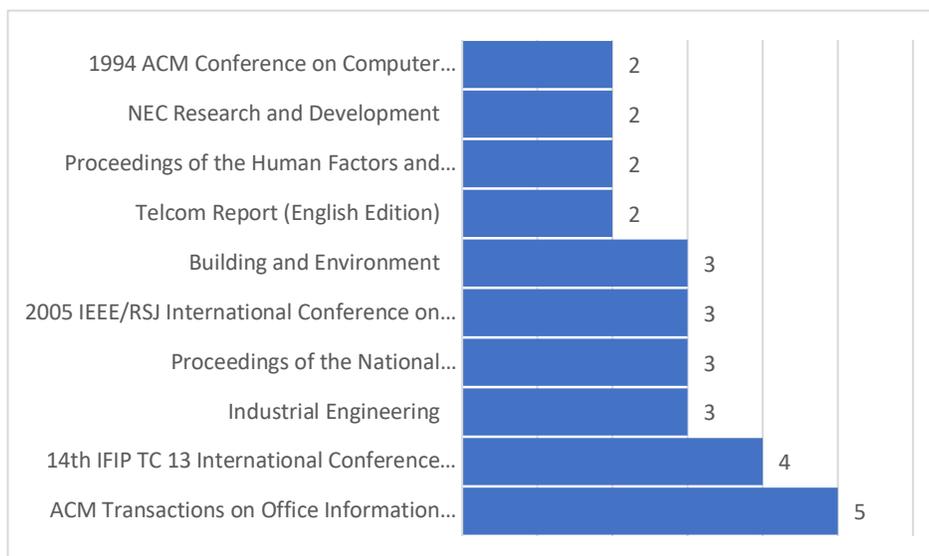

**Figure 6. Publication trends indicating the top journals and conference venues publishing studies on office work from the Compendex database.**

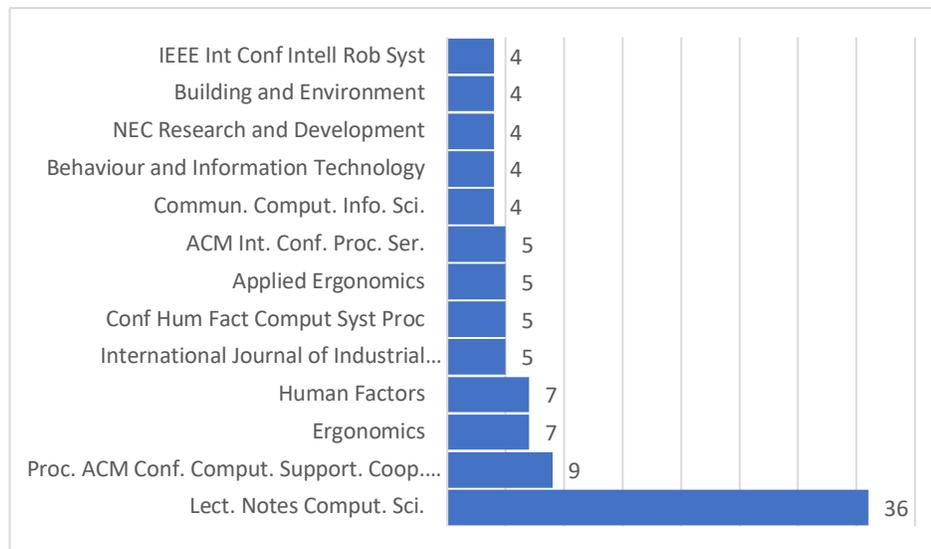

**Figure 7.** Publication trends indicating the top journals and conference venues publishing studies on office work from the Scopus database.

### 3.2 Knowledge Mapping and Visualization

#### 3.2.1. Top keywords

As indicated in tables 1 and 2, office automation, humans and human-computer interaction, and artificial intelligence were the top keywords that occurred in publications in both the Compendex and Scopus databases based on frequency of occurrence. Both databases also show publications with keywords related to buildings and office environments, which is a predominant theme in studies about office work.

**Table 1.** Top ten keywords occurring in research about office work, future of work in the Compendex database between 1970s- 2023.

| Keyword | Frequency of Occurrence | Total link strength |
|---|---|---|
| Office Automation | 62 | 85 |
| Human Computer Interaction | 49 | 92 |
| Artificial Intelligence | 44 | 81 |
| Automation | 30 | 66 |
| Office Buildings | 25 | 57 |
| Mobile Robots | 20 | 57 |
| Intelligent Robots | 18 | 60 |
| Interactive Computer Systems | 17 | 35 |
| Agricultural Robots | 15 | 42 |
| Ergonomics | 15 | 42 |

| Productivity | 14 | 23 |

The density visualizations in Figures 8 and 9 for the Compendex and Scopus databases also indicate that office automation, humans, and human-computer interaction, shown in yellow colors, occur more frequently and have a larger number of keywords connected to these keywords. Additionally, "ergonomics" features in both databases. Keywords denoting gender such as "male" and "female" occur among the top keywords in the Scopus database.

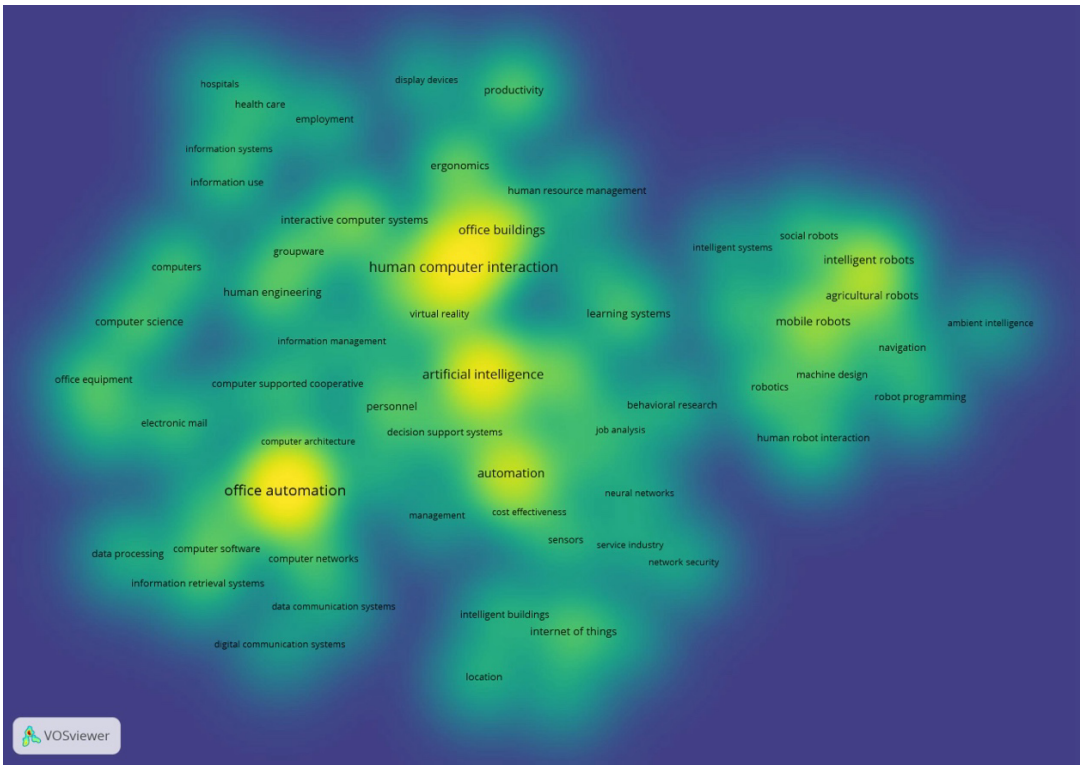

**Figure 8. Co-occurrence map showing density visualization indicating strength and significance of keywords from Compendex Database.**

Tables 1 and 2 indicate the total strength of the links between a given keyword and other keywords. Link strength represents the number of publications in which two keywords occur together, indicating the significance and relevance of the keyword to the topic under study. Based on the total link strengths, human computer interaction, office automation, and artificial intelligence have the highest strength in terms of links to other keywords in the Compendex database (Table 1), and humans, office workers, and females have the highest strength in terms of links to other keywords in the Scopus database (Table 2).

**Table 2. Top ten keywords occurring in research about office work, future of work in the Scopus database between 1970 - 2023.**

| Keyword | Frequency of Occurrence | Total link strength |
|---|---|---|
| Office Automation | 121 | 445 |
| Humans | 117 | 1200 |
| Human Computer Interaction | 116 | 709 |
| Automation | 105 | 556 |
| Office Environments | 94 | 231 |
| Office Workers | 79 | 856 |
| Artificial Intelligence | 73 | 215 |
| Female | 65 | 853 |
| Computers | 52 | 519 |
| Ergonomics | 49 | 383 |
| Male | 49 | 680 |

**Figure 9. Co-occurrence map showing density visualization indicating strength and significance of keywords from Scopus Database.**

3.2.2. Co-occurrence Maps

As can be seen in Figure 10, a bibliometric co-occurrence analysis of keywords from all the literature in the Compendex search indicates seven clusters of keywords. The seven clusters we identified in Compendex can be thematically categorized as containing literature on: (1) robotics; (2) human factors and computer-supported cooperative work; (3) networks, software, and data; (4) internet of things and intelligent buildings; (5) information systems in healthcare; (6) human resources, knowledge management, and decision support systems; and (7) theoretical facets of artificial intelligence. These clusters show that more recent studies have focused on social robotics and robotics in general, machine learning, internet of things, network security, hospital environments, ambient intelligence, and intelligent buildings. This has evolved from previous publications that focused more on office automation, computers, network security, office equipment, and information retrieval systems. In particular, the newer keywords focus more on machine learning, internet of things, and smart buildings. Office buildings seem to be the largest and most connected key term across all seven clusters of key terms that we identified. But there is some recent work on personnel, management, and human resources topics. Additionally, there are fairly recent articles on information systems used by office workers in healthcare systems.

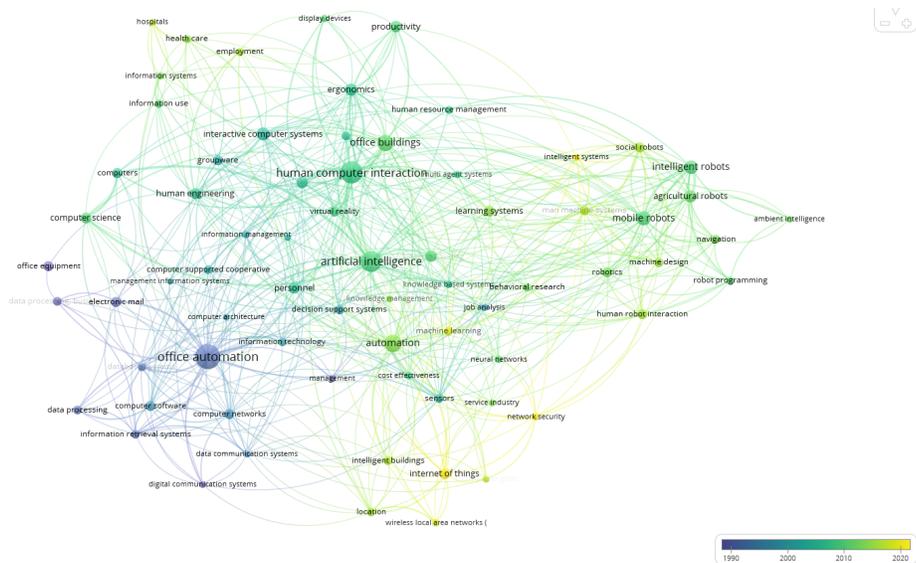

**Figure 10. Co-occurrence map showing keywords by year from Compendex Database.**

The co-occurrence analysis of keywords from the Scopus search indicates five separate clusters of keywords (figure 11). We can thematically categorize the five clusters identified from Scopus as follows: (1) office information systems and technologies; (2) worker characteristics; (3) ergonomics; (4) computers and cognition; and (5) humans and technology. Similar to Compendex results, more recent research focuses on robotics and office environments, but with a focus on worker characteristics and physical ergonomics. Earlier work focused more on office automation, computer networks and data processing, and cognitive components of office work.

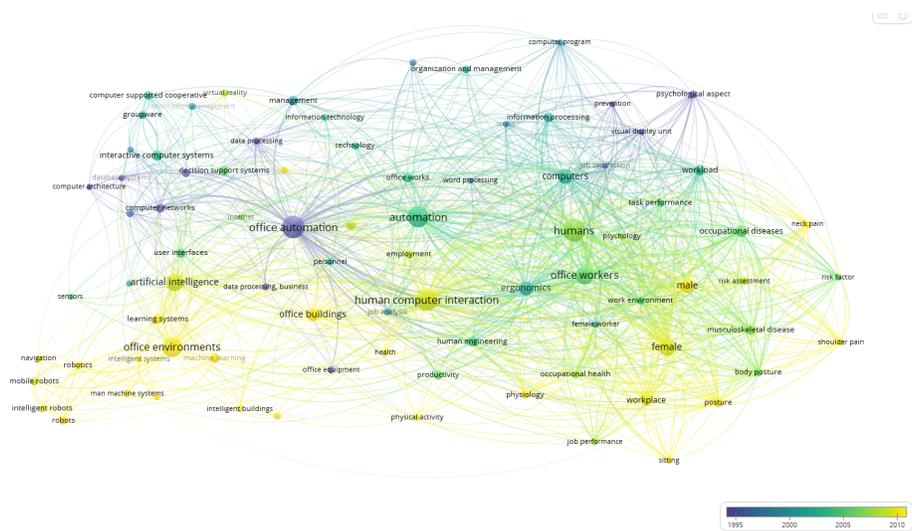

**Figure 11. Co-occurrence map showing keywords by year from Scopus Database.**

## 4. Discussion

The number of publications on office work-related topics in both Compendex and Scopus databases shows a spike in the 1980s, corresponding to the era of computerization when many research studies were conducted on office work (Figures 2 and 3). After the 1980s, the number of publications on this topic became stagnant until 2019–2020, after which the number of publications increased again. We believe COVID-19 and adaptations to remote work have increased the number of research studies related to work practices, tools and technologies, and management strategies needed for remote work. Artificial intelligence was still in development during this period, so the increase in the number of publications does not seem to correspond to the emergence of artificial intelligence tools such as ChatGPT or other similar developments.

The analysis of the types of publications across the two databases indicates that work has been published more often in conferences than in journals. We believe that the increased publication and presentation of work-in-progress at conferences suggests the possibility of ongoing development and resurgence in office work studies. But it may also be indicative of the focus on technology development in these studies, which may have been more suitable to publish at conferences for rapid feedback from research communities and from industry.

Findings also indicate that studies related to office work have been published in a wide range of journals and conferences, perhaps alluding to the interdisciplinary nature of problems in office work. Predominant disciplines include computer science, industrial engineering, computer-supported cooperative work, and human factors and ergonomics. This is not surprising given that the early foundational work and landmark studies [17], [18], [19], [20], [21], [22], [23], [24], [25] had their beginnings in the Computer Supported Cooperative Work (CSCW) and social science fields. Computerization also led to new ergonomics problems and subsequently to several research studies in posture, workload, and stress, which explains the human factors and ergonomics focus of these publications. Finally, both the development of software applications, robots, and hardware and human-computer interaction studies related to computerization and automation of office work are indicative of the disciplinary focus on computer science. Publications in the building and environment related fields and work in robotics point to emerging areas of focus on monitoring, sensing, and the use of robotics in office work. While the diverse disciplines represented in these journals point to the interdisciplinary nature of problems in office work, it is not evident whether the questions investigated in these studies are also multidisciplinary and cross-cutting. Our exploratory analysis indicates that these studies bound their research questions and gaps pertaining to a single discipline, but more in-depth studies on the gaps and research questions examined in these studies would help reveal the scope and disciplinary intersections of these research questions.

Our findings on the top keywords indicate that office automation, humans, human-computer interaction, and artificial intelligence occurred more frequently and had a higher link strength. This suggests that computerization, information systems, automation, and, more recently, artificial intelligence and their impact on people have become a focus of research in office work. In particular, how people interact with these information systems and how it impacts their productivity has been a longstanding research topic for investigation [26], [27], [28], [29], [30]. Similarly, we see an increase in emphasis on understanding the ergonomic impacts of computerization, including posture, workstation design, stress, and workload [31], [32], [33], [34], [35], [36], [37], with more technology development and implementation in office settings. Finally, we also note that, not surprisingly, many studies in office work have included a gender focus in understanding the critical office functions supported by female office workers and in highlighting the human resources implications [38], [39], [40]. This is not surprising given that a majority of office workers are women, and face unique challenges because of their gender.

The keyword clusters we identified from co-occurrence analysis are indicative of the main research themes in office work. They show progress in the office work domain and highlight emerging topics. We found seven keyword clusters in Compendex and five in Scopus.

Both Compendex and Scopus clusters indicate that intelligent buildings, robotics, activity monitoring and internet of things are emerging topics in the office work domain. In particular, we found smart and intelligent buildings and sensors to be a predominant theme. This indicates that a growing number of studies are using artificial intelligence, internet of things, and other advanced sensors and technologies to monitor built environments and assess interventions to increase comfort, improve productivity, and change physical activity behaviors in people to improve well-being [41], [42], [43], [44], [45], [46], [47], [48]. Researchers have also conducted research on environmental monitoring and sensing in the physical built infrastructure of office environments, with the aim of reducing energy consumption and associated costs [46], [49], [50], [51], [52], [53].

Much of the robotics research on this topic focuses on the development of programming and hardware related to the planning and execution of robot navigation [54], [55], [56], [57], [58], [59]. Office environments have primarily served as test settings for these new robot and technology developments. Some studies have also examined the evaluation of robots to aid office worker tasks like object fetching [60], [61], [62], [63]. Furthermore, the clustering of robotics, artificial intelligence, and other technological advances with older forms of information systems from the 1990s clearly demonstrates the evolution of office technology. The increase in studies on robot development, however, can imply future implementation and use in the office environment.

A separate cluster in Compendex on healthcare applications also indicates that domain specific investigations of office work are on the rise [64], [65], [66], [67]. This may be an important future direction given that the specifics of office work tasks, office worker characteristics, and skill requirements are vastly different across domains, pointing to a need for understanding training and professional development requirements with technological advances in specific domains.

Compared to the Compendex results, the clusters Scopus generated are more connected with other clusters of keywords from earlier research on the topic of office work. This shows how earlier work continues to inform recent studies on this topic and highlights the interdisciplinary nature of research that involves these topics. The two clusters related to ergonomics, worker characteristics, human performance, and safety indicate the types of human factors concerns that are more widely studied in office work. Given the sedentary nature of office work, most studies focus on posture, physical activity, and physical ergonomics [68], [69], [70], [71], [72].

Co-occurrence analysis of publications from Scopus indicates that while earlier work has focused more on office automation, computer networks and data processing, and cognitive components of office work, there has been less focus on cognitive components of office work in more recent studies involving artificial intelligence and other technological advances. This is somewhat surprising, given the importance of cognitive work in the context of artificial intelligence. We believe that work at the intersection of office work, cognitive work, and AI is emerging. Therefore, we suggest that future research examine the implications of AI on cognitive work.

### 4.1. Future Research Directions

The findings on the state of the art in office work research literature indicate that there are more studies on smart buildings, robotics, and technology and fewer on office workers and aspects important for worker development. We need to proactively understand how these technologies may actually support and augment office workers so that we can design technology to equitably benefit office workers. To achieve this goal, we need to conduct research studies investigating office workers, their activities, and their interactions concurrently or jointly with technology development studies to ensure that office worker considerations are not an afterthought. Hence, our findings point to an important need for more office worker-centered research studies investigating how they are adapting to the digital transformation, the implications for their professional development, and the impact on their well-being. Additionally, the office worker population is still predominantly female. Therefore, there is an urgent need for studies that explore the intersections between office worker gender, technology development, digital transformation, and labor displacement, and their impact on women's health and well-being in the workforce.

Our findings also point to a need for research on the implications of digital transformation in cognitive aspects of office work. In the earlier studies on office work, more emphasis was placed on physical ergonomics because of the need to understand workstation design and proper posture for effective computer-based work. Newer technological advances, such as artificial intelligence and robotic process automation, can shift how organizations design and implement processes and how office workers perform cognitive work activities such as data analysis and decision-making. This necessitates human factors engineering studies to investigate the impact of technology on organizational processes and, subsequently, on the cognitive activities of office workers. Such studies will help us understand how best to augment office worker activities that currently need support and will inform us of the new knowledge and skills needed for future office workers.

Our exploratory bibliometric findings indicate that while a majority of studies come from diverse disciplines, it is not clear whether the research questions and focus of the studies themselves are interdisciplinary. Our work identifies a need for interdisciplinary investigations on office work topics because of the cross-cutting problems and gaps in office work environments that warrant convergence to meaningfully impact changes for office workers. For example, technological advances such as artificial intelligence or changes in work modalities such as remote work or a four-day work week introduce significant changes to office worker tasks, how they perform their tasks, what skills, training, and education they need, and how human resources policies must be designed and implemented. A major technological change such as the introduction of AI can also introduce labor displacements, so it is equally important to examine equity and well-being challenges for office workers. To achieve optimal office worker-AI symbiosis in future work systems, we will need multidisciplinary teams with human factors engineers, experts in computer science and AI, diversity, equity and inclusion, economics, human resource management, labor relations, and public policy to comprehensively and systematically understand the implications of technological advances on office workers to inform the future of work system design.

In summary, scholarly literature demonstrates the significant implications for offices and office work based on the 1980s computerization era and hints at similar implications for the future. Given the ubiquitous nature of office work in all domains, its complexity, its importance in an organization's functioning and maintenance, productivity, and economics, and most importantly, the significant inequities that displacements in the office workforce will present to women, we will need to carefully and urgently study the future of office work in the AI age from a multidisciplinary lens.

## 5. Study Limitations

This bibliometrics analysis is limited to scholarly sources extracted from two databases that we determined to be the most relevant given the focus area of this study. We determined these databases and the search terms with the assistance of a research librarian. Bibliometrics analysis does not allow for evaluation of the quality of the publications or interpretation of the content in these publications [73]. Hence, these findings are indicative only of trends and the state of the art in the office work literature. Further work is needed to evaluate the in-depth content and the research questions examined in these publications.

## 6. Conclusions

Understanding the impact of technology, automation, and artificial intelligence on office and administrative support occupations is important for the benefits of digital transformation to fully support office workers. Our findings indicate that there are more studies on smart buildings, robotics, and technology and considerably fewer on office workers and the elements that impact their professional development. This points to the need for more office worker-centered research studies to be conducted concurrently or in conjunction with technology development studies so that office worker considerations are not an afterthought. Future research should focus on how these technologies can actually support and augment office workers so that technological advances are designed to equitably benefit office workers.